\documentclass[final,5p,times,twoside,twocolumn]{elsarticle}

\usepackage{graphicx,amssymb,amsmath}

\newcommand{\MS}{\overline{\mathrm{MS}}}
\newcommand{\MSb}{\MS}
\newcommand{\MM}{\mathrm{MM}}
\newcommand{\alphaMS}{\alpha^{\MS}_s}
\newcommand{\alphaMM}{\alpha^{\MM}_s}

\newcommand{\Eq}[1]{{Eq.~\eqref{#1}}}

\newcommand{\Tab}[1]{Tab.~\ref{#1}}
\newcommand{\Sec}[1]{Sec.~\ref{#1}}

\newcommand{\smallfrac}[2]{\tfrac{#1}{#2}}
\newcommand{\preprint}{
  \setlength{\unitlength}{1mm}{\hbox{\begin{picture}(0,0)
        \put(0,10){\mbox{\footnotesize%
            ADP-09-03/T681}}\end{picture}}}}

\biboptions{sort&compress}
\journal{Physics Letters B}

\begin{document}

\begin{frontmatter}
\title{\preprint
  The strong coupling and its running to four loops in a minimal MOM scheme} 

\author[cssm,darmstadt]{Lorenz von Smekal}
\author[cssm,york]{Kim Maltman}
\author[cssm]{Andr\'e Sternbeck}
\address[cssm]{CSSM, School of Chemistry \& Physics, The University of
  Adelaide, SA 5005, Australia}
\address[darmstadt]{Institut f\"ur Kernphysik,
  Technische Universit\"at
  Darmstadt, Schlossgartenstr.~9, D-64289 Darmstadt, Germany}
\address[york]{Department of Mathematics
  and Statistics, York University, Toronto, ON, M3J 1P3, Canada}

\date{August 5, 2009}

\begin{abstract}
  We introduce the minimal momentum subtraction (MiniMOM) scheme for
  QCD. Its definition allows the strong coupling to be fixed solely
  through a determination of the gluon and ghost propagators. In Landau
  gauge this scheme has been implicit in the early studies of these
  propagators, especially in relation to their non-perturbative
  behaviour in the infrared and the associated infrared
  fixed-point. Here we concentrate on its perturbative use. We give the
  explicit perturbative definition of the scheme and the relation of its
  $\beta$-function and running coupling to the $\MSb $ scheme up to
  4-loop order in general covariant gauges. We also demonstrate, by
  considering a selection of $N_f=3$ examples, that the apparent
  convergence of the relevant perturbative series can in some (though
  not all) cases be significantly improved by re-expanding the $\MSb$
  coupling version of this series in terms of the MiniMOM coupling,
  making the MiniMOM coupling also of potential interest in certain
  phenomenological applications.
\end{abstract}


\end{frontmatter}

\section{Introduction}
\label{intro}

The coupling constant of the strong interaction, $\alpha_s$, is one of
the fundamental parameters of the Standard Model. It is specified by
its value in a particular renormalisation scheme at a chosen
reference scale, $\mu$, conventionally taken to be the $\MSb$
scheme at the $N_f=5$ reference scale $\mu =M_Z$.
A recent assessment of experimental results
yields $\alpha^{\MSb}_s(M_Z)=0.1189(10)$ \cite{Bethke:2006ac}. This result,
which is little changed if more recent experimental results are taken into 
account (see, e.g., \cite{Maltman:2008bx} and references therein), is in
excellent agreement with two recent, slightly different, lattice 
determinations \cite{Maltman:2008bx,Davies:2008sw}
based on lattice perturbation theory analyses of short-distance-sensitive
lattice observables computed using the MILC $N_f=2+1$ configurations.

Other schemes than the $\MSb$ scheme are of course also possible. For
example it has been proposed in
\cite{vonSmekal:1997isvonSmekal:1997vx} that a particular 
product of dimensionless gluon and ghost dressing functions, $Z$ and
$G$, in the Landau gauge can be used to define a non-perturbative
running coupling via 
\begin{equation} 
 \alpha^{\MM}_s(p^2) \, = \, \frac{g^2}{4\pi} Z(p^2) G^2(p^2) \; ,
 \label{alpha_minimom}
\end{equation}
where $g^2 \equiv g^2(\mu)$ is the renormalised coupling at scale
$\mu$ and the renormalisation condition,  
\begin{equation} 
  Z(p^2) G^2(p^2) \big\vert_{p^2 = \mu^2} = \, 1 \; ,
\label{np-ren}
\end{equation}
is assumed. The definition \eqref{alpha_minimom} is particularly
convenient since it allows the coupling and hence $\Lambda^{\MSb}$ to be
determined by measuring two-point functions on a lattice. Our first
steps towards such a determination for $N_f=0,2$ were presented at the
2007 Lattice Conference \cite{Sternbeck:2007br}. Further updates of
these preliminary results were reported last year
\cite{vonSmekal:2008ma,Sternbeck:2008au}. This project, which is
ongoing, has the potential
to provide an independent precision determination of $\alpha_s$ from
lattice simulations at purely perturbative scales, typically between
$10 $ and $100\, \textrm{GeV}$~\cite{Sternbeck:2009la}.\footnote{The
  QCD scale parameter of the underlying scheme (the MiniMOM scheme,
  see below) is roughly 450~MeV for  $N_f=0$, or 430~MeV for $N_f=2$. 
  Non-perturbative contributions to the gluon and ghost dressing
  functions are at least of the order
  $(\Lambda_{\mathrm{QCD}}/\mu)^2$. They are negligible at scales
  $\mu$ above  $10~\mathrm{GeV}$.}
As an important supplement we specify here the details of the
renormalisation scheme underlying the coupling \eqref{alpha_minimom}
for QCD in general covariant gauges. We also provide the explicit
4-loop $\beta$ function for $\alphaMM$ in all such gauges. This
information will be important to the previously mentioned lattice
analysis and to quantifying the truncation error on the resulting
$\alpha_s$ determination.

The product in \eqref{alpha_minimom} is dimensionless and renormalisation group
invariant, and it reduces to the running coupling of a perturbative
momentum subtraction scheme (MOM) at large~$p^2$. This makes it a 
suitable candidate for a non-perturbative extension, though such
extensions are, of course, not unique. The underlying
renormalisation condition \eqref{np-ren} respects infrared scaling 
(with $0.5 < \kappa < 1$ \cite{vonSmekal:1997isvonSmekal:1997vx}),
\begin{equation}
   \label{infrared-gh_gl}
  Z(p^2) \, \sim\, (p^2/\Lambda^2_{\mbox{\tiny QCD}})^{2\kappa} \; , \;\; 
  \; \mbox{and} \;\;  G(p^2) \, \sim \, 
  (p^2/\Lambda_{\mbox{\tiny QCD}}^2)^{-\kappa} \; ,
\end{equation}
for $p^2 \to 0$, as predicted by a variety of
functional continuum methods for Landau gauge QCD including studies of
Dyson-Schwinger Equations (DSEs) \cite{Lerche:2002ep}, Stochastic
Quantisation \cite{Zwanziger:2001kw}, and of the Functional
Renormalisation Group Equations (FRGEs)~\cite{Pawlowski:2003hq}. This
conformal infrared behaviour of the purely gluonic correlations in
Landau gauge QCD is consistent with the conditions for confinement in
local quantum field theory
\cite{Alkofer:2000wg,Alkofer:2001iwAlkofer:2000mz}, but 
it is yet to be observed in lattice simulations.\footnote{In order to
  do that one needs a proper non-perturbative definition of BRST symmetry on
  the lattice which is possible in principle, but not 
    realised in present lattice implementations of Landau gauge.
    For a recent discussion, see \cite{vonSmekal:2008ws}.}

If the infrared scaling behaviour \eqref{infrared-gh_gl} is realised,
the renormalisation condition (\ref{np-ren}) holds beyond
perturbation theory and can be imposed at any (space-like) subtraction
point $p^2=\mu^2 \ge 0$. This is because the running coupling defined
by \eqref{alpha_minimom} then approaches a finite infrared fixed-point,
$\alpha^{\MM}_S \to \alpha_c$ for $p^2 \to 0$, with $\alpha_c \approx
8.9/N_c$ obtained under a mild regularity assumption 
on the ghost-gluon vertex \cite{Lerche:2002ep}. For the purposes of
this Letter it suffices, however, that the coupling
\eqref{alpha_minimom}, is well-defined perturbatively,
independent of the infrared scaling behaviour in \eqref{infrared-gh_gl}.

A special feature of Landau gauge which underlies the definition
\eqref{alpha_minimom} is the non-renormalisation of the ghost-gluon vertex
\cite{Taylor:1971ff}. 
The possibility of taking advantage of this non-renormalisation
has been criticised in the past
\cite{Boucaud:2005ce,Boucaud:2008ji} on the grounds that the
Landau-gauge ghost-gluon vertex acquires a momentum dependence in
all common MOM schemes, already at one-loop level
\cite{Celmaster:1979km}, despite the absence of 
ultraviolet divergences in this vertex in Landau gauge. There is, however, no
contradiction here at all, as will be explained in the next section. 
The basic reason is that the non-renormalisation of the ghost-gluon vertex
ensures the existence of a scheme for which the notion of 
(\ref{alpha_minimom}) as a running coupling makes sense and which this
running coupling implicitly defines, without the need to use an asymmetric
momentum scheme \cite{Boucaud:2008gn}. We call it the
MiniMOM scheme for reasons that will become clear below. A useful
feature of the MiniMOM scheme is that it can be related to the $\MSb$
scheme at four-loops \cite{vonSmekal:2008ma,Sternbeck:2008au}
without the need to compute vertices to three loops in perturbation theory. 

After providing more details on the MiniMOM coupling and its relation
to the $\MSb$ coupling below, we consider, in \Sec{sec:applics}, 
the $\MSb$ and MiniMOM versions of the perturbative series entering
a selection of  phenomenological applications, demonstrating
that, in some cases, the apparent convergence of the series
is much improved by the use of the MiniMOM coupling.
Our conclusions and outlook are given in \Sec{sec:conclusion}.

\section{The minimal MOM scheme}
\label{sec:MiniMOMscheme}

Some confusion in the literature concerning the running coupling
of Eq.~(\ref{alpha_minimom}) arose in relation to the
misconception that this definition rests on the non-renormalisation
of the ghost-gluon vertex in Landau gauge \cite{Taylor:1971ff}. This
definition is not, however, based on requiring that the ghost-gluon
vertex reduce to the tree-level one at a symmetric subtraction point
$k^2= p^2 = q^2 = \mu^2$.  In particular, the scheme underlying
(\ref{alpha_minimom}) is not the MOMh scheme of Ref.~\cite{Chetyrkin:2000fd},
but is, instead, a minimal MOM (MiniMOM) scheme, which is defined as follows:

As in every MOM scheme, the gluon and ghost renormalisation constants
$Z_3$ and $\widetilde Z_3$ are defined by requiring 
\begin{equation}
    Z(p^2)\big|_{p^2=\mu^2} = 1 \quad \mbox{and} \quad
    G(p^2)\big|_{p^2=\mu^2} = 1 \; ,
\label{pt-ren}
\end{equation}
the perturbative realisation of (\ref{np-ren}) valid for $\mu^2 \gg
\Lambda^2$ where $\Lambda $ is the scale parameter of the scheme. 
$Z(p^2) $ and $G(p^2)$ are the dressing functions 
in the renormalised gluon and ghost propagators of 
 Landau-gauge QCD, which are in (Euclidean) momentum space of the form
\begin{align}
  \label{eq:gl_dress}
  D^{ab}_{\mu\nu}(p) \,&=\, 
  \delta^{ab}\left(\delta^{\mu\nu}-\frac{p_{\mu}p_{\nu}}{p^2}\right) 
  \frac{Z(p^2)}{p^2}\\
\intertext{and}
\label{eq:gh_dress}
  D_G^{ab}(p)   \,&=\,  -\delta^{ab}\;\frac{G(p^2)}{p^2} \; ,
\end{align}
respectively. Instead of imposing additional analogous renormalisation
conditions on vertex functions, requiring certain vertex structures
to equal their tree-level counter parts at some symmetric or
asymmetric subtraction point, we here supplement (\ref{pt-ren}) by the
further condition
\begin{equation} 
\widetilde Z_1 \, = \, \widetilde Z_1^{\MSb} \; ,   \label{vert-MM}
\end{equation}
where $\widetilde Z_1$ is the renormalisation constant of the
ghost-gluon vertex, whose momentum dependence is thus the same as in
the minimal subtraction schemes. The renormalisation constants for the
three and four gluon vertex functions are then determined by the
Slavnov-Taylor identities as usual,
\begin{equation} 
  Z_1 \, = \, \frac{Z_3}{\widetilde Z_3} \,  \widetilde Z_1 \quad
  \textrm{and}  \quad Z_4 \, = \, \frac{Z_3}{\widetilde Z_3^2} \,
\widetilde Z_1^2 \; .
\end{equation}
The extension to quarks is straightforward, and this defines the
MiniMOM scheme: momentum subtraction for the gluon, ghost and quark
propagators, and minimal ($\MSb$) subtraction for the ghost-gluon vertex
(\ref{vert-MM})  
together with the Slavnov-Taylor identities to fix the remaining
vertices, including the quark-gluon vertex. Note that the renormalisation
constants for these remaining vertices then differ from those in the $\MSb$
scheme by ratios of the propagator (field) renormalisation constants in the
MiniMOM ($\MM$) and $\MSb$ schemes, for example,
\begin{equation} 
  Z_1^\MM  = \frac{Z_3^{\MM}}{Z_3^{\MSb}}\, \frac{\widetilde
  Z_3^{\MSb}}{\widetilde Z_3^{\MM}} \,  Z_1^{\MSb} \quad \mbox{and} 
 \quad Z_4^\MM = \frac{Z_3^\MM}{Z_3^{\MSb}}\,
\left(\frac{\widetilde Z_3^{\MSb}}{\widetilde Z_3^\MM}\right)^2 Z_4^{\MSb}\; .
\end{equation}
Likewise, from the general definition of the renormalised coupling constant,
\begin{equation} 
  \alpha_s(\mu)  \,=\,  \frac{g^2(\mu)}{4\pi} \,= \,  \frac{ Z_3
  \widetilde Z_3^2}{\widetilde Z_1^2} \,
  \frac{g^2_\mathrm{bare}}{4\pi} \; ,  
\end{equation}
the MiniMOM and $\MSb$ scheme couplings, with the definition 
$\widetilde Z_1^\MM \equiv  \widetilde Z_1^{\MSb}  $ in (\ref{vert-MM}), 
are related by
\begin{equation} 
\alpha_s^\MM(\mu)  \, = \,  \frac{ Z_3^\MM }{
  Z_3^{\MSb} } \,  \left( \frac{\widetilde Z_3^{\MM}}{\widetilde
  Z_3^{\MSb}} \right)^2   \,   \alpha_s^{\MSb}(\mu) \; .
\end{equation}
Note that for this conversion we only need to know the perturbative
expansions of the ghost and gluon propagators but not of any vertex
structures. Rather, with the MOM renormalisation conditions
for these propagators (\ref{pt-ren}), or their non-perturbative
extension (\ref{np-ren}) for that matter,  we simply obtain,
\begin{equation} 
  \frac{\alpha_s^\MM(\mu) }{  \alpha_s^{\MSb}(\mu) } \, = \, 
  Z(\mu^2)^{\MSb} G^2(\mu^2)^{\MSb} \; , 
\label{ZG2-MSb}
\end{equation}
from the gluon and ghost dressing functions evaluated at $p^2 = \mu^2$
in the $\MSb$ scheme. Alternatively, we can relate the MiniMOM
coupling to the MOMh scheme just as easily via
\begin{equation}
  \alpha_s^\MM(\mu)  \, = \,  \left( \frac{\widetilde
  Z_1^\mathrm{MOMh}}{\widetilde 
  Z_1^{\MSb}} \right)^2   \,   \alpha_s^{\mathrm{MOMh}}(\mu) \; .
  \label{conv-MOMh}
\end{equation}
All these conversion identities are valid for arbitrary
linear-covariant gauges and not restricted to Landau gauge. The
special feature of Landau gauge is that there $\widetilde Z_1^{\MSb} =
\widetilde Z_1^\MM = 1$. Because the Landau gauge ghost-gluon vertex
trivially reduces to its tree-level form when one of the ghost momenta
is set to zero, the MiniMOM scheme then agrees with the asymmetric
$\widetilde{\mathrm{MOM}}\mathrm{h}$ scheme of
Ref.~\cite{Chetyrkin:2000dq} (called the $T$ scheme in
Ref.~\cite{Boucaud:2008gn}) which is defined from renormalising
precisely this vertex structure.  The MiniMOM scheme is defined,
however, so as to not require knowledge of any vertex structure beyond
the $\MSb$ scheme contributions as determined by their ultra-violet
divergences. This makes it particularly useful for a lattice
  determination  of $\alpha_s$ from the perturbative behaviour of QCD
  Green's functions as described in
  Refs.~\cite{Sternbeck:2007br,vonSmekal:2008ma,Sternbeck:2008au}.  

For early 2 and 3-loop calculations in Feynman gauge, see
\cite{Tarasov:1976efTarasov:1980kxTarasov:1980au}. The complete
2-loop results for general gauge parameters are given in
\cite{Davydychev:1997vhDavydychev:1998kj}. To obtain the 4-loop version 
of the conversion between $\alpha_s^{\MSb}$ and $\alpha_s^\MM $ from
Eq.~(\ref{ZG2-MSb}), we use the 3-loop expressions for the gluon and ghost 
self-energies found in Appendix~C of \cite{Chetyrkin:2000dq} which, 
with $a\equiv \alpha_s^{\MSb}/\pi$, for $N_f $ massless quarks yields
\begin{equation}
  \alpha_s^\MM /\alpha_s^{\MSb} \,=\, 1 \, + \, D_1 \, a \, +
  D_2 \, a^2 \, + \, D_3 \, a^3 \, + \, \mathcal O(a^4) \; , \label{MM-MSb-4L}
\end{equation}
where, with $\zeta_3=1.2020569$, $\zeta_5=1.0369278$,
and $C_F=(N_c^2-1)/2N_c$,
\begin{subequations}
  \label{eqs:Ds}
{\allowdisplaybreaks
\begin{align} 
   D_1 &= d_{10} \, + \, d_{11} \, N_f \; ,  \label{ZG2-MSb-expl-1}\\ 
   & d_{10} = \Big[\smallfrac{169}{144} + \smallfrac{1}{8} \, \xi +
   \smallfrac{1}{16}\, \xi^2 \Big] N_c \; , \;\; d_{11} =  -
   \smallfrac{5}{18}\; .
  \nonumber \\[2pt]
   D_2 &= d_{20} \, + \, d_{21} \, N_f \, + \, d_{22} \, N_f^2 \;
   ,  \label{ZG2-MSb-expl-2} \\ 
   & d_{20} = \Big[\smallfrac{76063}{20736} - \smallfrac{39}{128}  \zeta_3
+ \Big( \smallfrac{269}{2304} + \smallfrac{11}{64} \zeta_3 \Big)\, \xi\,
   + \nonumber\\
& \hskip .9cm 
\Big(\smallfrac{35}{256} - \smallfrac{3}{128} \zeta_3 \Big) \, \xi^2 +
   \smallfrac{5}{256}\,  \xi^3 \Big]  N_c^2\; ,  \nonumber\\
  & d_{21} = - \Big[ \smallfrac{55}{96} - \smallfrac{1}{2}
\zeta_3 \Big] C_f  - \Big[ \smallfrac{1583}{1296} + \smallfrac{1}{4}
\zeta_3 + \smallfrac{5}{144} \, \xi \Big] N_c \; ,  \nonumber \\
& d_{22} = \smallfrac{25}{324} \; .  \nonumber\\[2pt]
D_3 & =   d_{30} \, + \, d_{31} \, N_f \, + \, d_{32} \, N_f^2  \, +
   \, d_{33} \, N_f^3 \; ,  \label{ZG2-MSb-expl-3} \\
& d_{30} = \Big[ \smallfrac{42074947}{2985984} - \smallfrac{20225}{9216}
  \zeta_3 - \smallfrac{7805}{12288} \zeta_5 +  \nonumber\\
& \hskip .9cm \Big( \smallfrac{17743}{41472} + \smallfrac{10335}{9216}
   \zeta_3 - \smallfrac{295}{1024} \zeta_5 \Big) \, \xi \, +  \nonumber\\
& \hskip .9cm \Big( \smallfrac{16235}{36864}-\smallfrac{71}{1536}
   \zeta_3  - \smallfrac{175}{6144} \zeta_5 \Big) \, \xi^2 \, + 
\nonumber\\
& \hskip .9cm \Big( \smallfrac{1207}{12288}-\smallfrac{61}{3072}
\zeta_3 - \smallfrac{5}{3072}  \zeta_5  \Big) \, \xi^3 \, +
\nonumber\\ 
& \hskip .9cm \Big( \smallfrac{169}{12288}-\smallfrac{11}{3072}
\zeta_3 + \smallfrac{35}{12288}  \zeta_5  \Big) \, \xi^4 \Big] N_c^3 
\; ,   \nonumber\\ 
& d_{31} = \Big[
 - \smallfrac{202997}{31104} - \smallfrac{217}{288} \zeta_3 
 + \smallfrac{5}{12} \zeta_5 - \nonumber\\
& \hskip .9cm  \Big( \smallfrac{505}{5184} + \smallfrac{17}{48}
 \zeta_3 \Big) \, \xi  \, -  \Big( \smallfrac{497}{9216} - \smallfrac{1}{128}
 \zeta_3 \Big) \, \xi^2   \Big] N_c^2  -  \nonumber\\
& \hskip .9cm 
 \Big[\smallfrac{41243}{10368} - \smallfrac{41}{16} \zeta_3 
 -  \smallfrac{5}{8} \zeta_5 + \Big( \smallfrac{55}{768} -
 \smallfrac{1}{16} \zeta_3 \Big) \, \xi \, \Big] C_f N_c  + \nonumber\\ 
& \hskip .9cm 
 \Big[ \smallfrac{143}{576} + \smallfrac{37}{48} \zeta_3  -
   \smallfrac{5}{4} \zeta_5 \Big] C_f^2  \; ,  \nonumber\\
& d_{32} =  \Big[ 
   \smallfrac{47965}{62208} +\smallfrac{37}{144} \zeta_3 +
   \Big(\smallfrac{7}{2592} + \smallfrac{1}{36} \zeta_3 \Big)
   \, \xi \, \Big] N_c + \nonumber\\ 
& \hskip .9cm    \Big[ \smallfrac{7001}{10368}
 - \smallfrac{13}{24} \zeta_3 \Big]
  C_f \; , \;\;  
  d_{33} =   - \smallfrac{125}{5832} \; . \nonumber
\end{align}}
\end{subequations}
This conversion depends on the gauge parameter $\xi \equiv \xi^{\MSb}$
as in every other momentum subtraction scheme, though
in the MiniMOM scheme, this 
dependence is comparatively weak. At one-loop
level, for example, the coefficient of the leading $\xi$
dependence around $\xi = 0$ in the above conversion is 3 times smaller
than that of the asymmetric $\widetilde{\mathrm{MOM}}\mathrm{h}$
scheme (which coincides with the MiniMOM scheme at $\xi=0$).

For $\xi = 0$ the same conversion can be obtained from
the product of the scheme-invariant propagators given for Landau
gauge in Sec.~4 of \cite{Chetyrkin:2004mf}. We verified the general
result \eqref{eqs:Ds} for $N_c=3$ and $N_f = 0, \, 3$
and $6$ from the explicit expressions given there.
Another check of our conversion, 
up to and including $O(a^2)$, can be obtained using
\Eq{conv-MOMh} in Landau gauge, where $\widetilde Z_1^{\MSb} = 1 $, 
with the 2-loop expression for $ \widetilde Z_1^\mathrm{MOMh}(\mu)  =  \widetilde
\Gamma^{-1}(\mu^2) $ in the symmetric $\mathrm{MOMh}$ scheme 
from \cite{Chetyrkin:2000fd}, together with the 3-loop conversion from
$\alpha_s^{\MSb}$ to $\alpha_s^\mathrm{MOMh}$ as given there.\footnote{There  
  appears to be a typo in Table 2 of \cite{Chetyrkin:2000fd}, the
  entry for the 2-loop coefficient of $\widetilde \Gamma $
  proportional to $N_c N_f$ should probably read $ -0.115(2)$ instead
  of $- 0.151(2)$.}    

The numerical values of the $D_k $ in Landau gauge
are given explicitly, for $N_c=3$ and $N_f = 0,\, 2,$ and $3$,
in \Tab{tab:Ck_nf}. As an illustration of the weakness of the dependence
on $\xi$, the corresponding $N_c=3$, $N_f=3$ results are
\begin{align} 
 D_1 =& \ 2.6875+0.3750 \xi+0.1875 \xi^2 , \\
 D_2 =& \ 16.8264+2.5977 \xi+0.9769 \xi^2 +0.1758 \xi^3 , \nonumber\\
 D_3 =& \ 127.6687+26.7739 \xi+8.3907 \xi^2+1.9621 \xi^3+0.3349 \xi^4,
\nonumber 
\end{align}
with similar, slightly less $\xi$-dependent, results 
for $N_f=0,2$.

\begin{table}[bt]
  \label{tab:Bxi}
  \centering
  \caption{The numerical values of the $D_k $ in the 4-loop
     relation between $\alpha_s^\MM $ and $\alpha_s^{\MSb} $ in Landau
     gauge, Eqs.~\eqref{MM-MSb-4L} and \eqref{eqs:Ds}
     with $\xi = 0$, for $N_c=3 $ and $N_f = 0,\, 2,$ and $3$.}
  \label{tab:Ck_nf}\small
  \begin{tabular*}{\linewidth}{@{\extracolsep{\fill}}r@{\qquad}rrr}
    \hline\rule{0pt}{2ex} 
    & $N_f=0$ & $N_f=2$ & $N_f=3$ \\*[0.5ex]
    \hline\rule{0pt}{2ex}
    $D_1$ &   3.52083333    &   2.96527778    &   2.6875 \\
    $D_2$ &  29.71718945    &  20.96900712    &  16.82639745 \\
    $D_3$ & 291.4436449     & 175.9308786     & 127.6686773 \\
    \hline
  \end{tabular*}
\end{table}

In order to compare actual values of $\alpha_s$ in the MiniMOM
  scheme (at $\xi =  0$) to the corresponding ones
  in the $\MSb$ scheme we give two important examples.
  First, at the mass of the $Z$ boson, $m_{Z} = 91.2$ GeV with
  $\alpha_s^{\MSb}(m_{Z}^2) = 0.1189 $  
  \cite{Bethke:2006ac}, from Eq.~(\ref{MM-MSb-4L}) we obtain 
\begin{equation} 
     \alpha_s^{\MM}(m_{Z}^2) = 1.096 \;  \alpha_s^{\MSb}(m_{Z}^2) \,
     ,  \label{eq:comp_Z} 
\end{equation}
for $N_f = 5$, while at the mass of the $\tau $ lepton, $m_\tau = 1.777
$ GeV,
\begin{equation} 
     \alpha_s^{\MM}(m_\tau^2) = 1.59 \;  \alpha_s^{\MSb}(m_\tau^2) \,
     , \label{eq:comp_tau}
\end{equation}
where, to be specific, we have used $\alphaMS(m^2_\tau) = 0.322$, the
value obtained by running $\alphaMS(m_{Z}^2) = 0.1189$ down to the
$N_f=3$ scale $m^2_\tau$ using the standard 4-loop running
\cite{Chetyrkin:1997sg}.

  We conclude this section with a few comments on quark mass effects. As
  in every off-shell subtraction scheme, the running coupling and beta
  function of the MiniMOM scheme in principle depend on the quark
  masses. This is evident from its relation to the (mass independent)
  $\MSb $ scheme, Eq.~(\ref{ZG2-MSb}), in which the $\MSb $ scheme
  gluon and ghost dressing functions will depend on the masses in the
  quark loops. These have not been included and our conversion
  formulas are therefore strictly speaking valid only for $N_f$
  massless quarks. To fully account for finite quark masses at this
  level one would need the corresponding 3-loop expressions for the
  gluon and ghost dressing functions with massive quark loops which
  have not been calculated to our knowledge as yet. 

  We can estimate the leading quark
  mass effects, however, which will affect the conversion formula at the
  2-loop level. These are obtained from the 1-loop vacuum polarisation with
  massive fermions in the gluon self-energy, and they lead to an
  increase of $D_1$ (via $d_{11}$) as compared to the massless
  case. For each quark flavour with mass $m_q$ one then separately
  obtains a transcendental function $d_{11}(y)$ of $y \equiv
  m_q^2/\mu^2$ which approaches $-5/18$ for $y \to 0$, i.e., for
  $\mu^2 \gg m_q^2$. Using the current upper limits from the Particle
  Data Group for the average up/down mass of $5 $ MeV and the strange
  mass of $130$ MeV as commonly given at $\mu = 2 $ GeV, we observe a
  maximum increase in $D_1$ by 0.15\% at $\mu = 2 $ GeV as compared to
  the massless $N_f = 3$ value given in \Tab{tab:Ck_nf}. At $\mu = 1$
  GeV, with correspondingly larger light quark masses, the same upper
  bound for the increase in $D_1$ reaches 1\%. At larger scales the
  effect rapidly decreases. In particular, the explicit comparisons in
  Eqs.~(\ref{eq:comp_Z}) and (\ref{eq:comp_tau}) remain unaffected by
  the corresponding changes in $D_1$ (note that even with the charm
  and bottom quark masses included, the increase in $D_1$ will be less
  than 0.1\% at $\mu=m_Z$ as compared to the $N_f = 5$ massless
  flavour value there).       

  The most noticeable quark mass effects will of course occur right at the
  decoupling scales $\mu = m_q$. At the charm threshold, for example,
  with $\mu = m_c = 1.27 $ GeV, the quark mass contributions to the
  vacuum polarisation lead to an increase in $D_1$ by around 13\% as
  compared to the massless $N_f=4$ value (at $\mu=2$ GeV this increase
  is reduced by a factor of 2 already). Considerable charm-quark mass
  effects should therefore be expected when converting the 4 flavour MiniMOM
  coupling to the $\MSb $ coupling in the phenomenoligically
  interesting range between 1 and 4 GeV. Without a more detailed knowledge
  of these effects an $N_f=4$ conversion
  would therefore not be advisable. Here we use the
  MiniMOM to $\MSb$ conversion up to at most $N_f=3$ for which the quark
  mass effects are very small. For now, matching to the $N_f=4$ and $5$ regimes
  should always be done after the conversion, for the $\MSb $
  coupling in the usual way. 

  There are no charm and bottom quarks in the  
  lattice determinations of the MiniMOM coupling which are presently
  restricted to $N_f = 0$ and $2$ and which will be extended to $2+1$
  light flavours in due course. At the relevant high momentum scales
  quark mass effects should then be completely negligible in the
  conversion to $\alpha_{\MSb}$. In addition, it is always a
  possibility to remove any small residual light-quark mass effects by
  extrapolation, if necessary.

\section{Beta-function coefficients of the MiniMOM coupling} 
\label{sec:the_MM_betafunction}

The running of the coupling constant as the scale, $\mu$, changes is
controlled by the (scheme-dependent) $\beta$ function which, at small
couplings, is defined by
\begin{equation}
  \mu^2{\frac{da(\mu^2)}{d\mu^2}} = \beta(a) := -\sum_{i=0}\beta_i a^{i+2}\, ,
\end{equation}
where $a\equiv\alpha_s/\pi$. 
The $\beta$ function in the $\MSb$ scheme is known to
4-loop order, the expressions for the corresponding
coefficients $\beta_k^{\MS}$, $k=0,\cdots ,3$, 
for general $N_c$ and $N_f$, being given in 
Refs.~\cite{vanRitbergen:1997vaCzakon:2004bu}. These results,
together with the relation between the $\MSb$ and MiniMOM coupling
given in Eqs.~(\ref{MM-MSb-4L}) and \eqref{eqs:Ds},
and the 3-loop version of the expression
$\mu^2d\xi/d\mu^2=\xi\gamma_3$,
for the running of the renormalised $\MS$ gauge parameter, $\xi$, 
with $\gamma_3$ the gluon anomalous dimension, the 3-loop expression 
for which can be found in Appendix~D of Ref.~\cite{Chetyrkin:2000dq}, 
allow us to obtain the $\beta$ function coefficients in the MiniMOM 
scheme to 4-loop order. For general $N_c$ and $N_f$, we find
\begin{subequations}\label{eqs:betaCoeff}
{
\begin{align}
\beta^{\MM}_0&={\smallfrac{1}{4}}
\left[ {\smallfrac {11}{3}}\, N_c -{\smallfrac{2}{3}}\,N_f \right]\;, 
\\*[1.5ex]
\beta_1^{\MM}&=\smallfrac {1}{8}\left[
{\smallfrac {17}{3}} N_c^{2}-{\smallfrac {5}{3}}N_f N_c 
-C_f N_f \right] + B^{\xi}_{10} + B^{\xi}_{11}N_f\;,\\*[1.5ex]
\beta_2^{\MM}&=
 \left[ {\smallfrac {9655}{4608}}
-{\smallfrac {143}{512}}\, \zeta_3 \right] \, N_c^{3} -
\left[ {\smallfrac {2009}{2304}}
+{\smallfrac {137}{768}}\, \zeta_3\right] \, N_f N_c^{2}
\\
&\quad
+\left[ \smallfrac {23}{384}+{\smallfrac{1}{24}}\, \zeta_3
\right]\, N_f^{2} N_c 
-\left[ {\smallfrac {641}{1152}}
-{\smallfrac {11}{24}}\,\zeta_3 
\right]\, C_f N_f N_c \nonumber\\
&\quad
+{\smallfrac {1}{64}} C_f^{2}N_f
+\left[ {\smallfrac {23}{288}}-{\smallfrac{1}{12}}\,\zeta_3 
\right]\, C_f N_f^{2} \nonumber \\
&\quad
+B^{\xi}_{20}+B^{\xi}_{21}N_f+B^{\xi}_{22}N_f^2\;,
\nonumber\\*[1.5ex]
\beta_3^{\MM}&=
\left[ {\smallfrac {1381429}{165888}} 
-{\smallfrac {225335}{110592}}\, \zeta_3 
-{\smallfrac {85855}{73728}}\,\zeta_5 \right] N_c^{4}
\\
&\quad
+\left[
-{\smallfrac {244549}{55296}} 
+{\smallfrac {3395}{18432}}\,\zeta_3 
+{\smallfrac {35965}{36864}}\,\zeta_5
\right] N_f N_c^{3}\nonumber\\
&\quad
+\left[ -{\smallfrac {5}{96}}
+{\smallfrac {11}{8}}\,\zeta_3
-\left( {\smallfrac {60685}{18432}} 
-{\smallfrac {85}{64}}\,\zeta_3 
-{\smallfrac {55}{48}}\,\zeta_5\right) C_f  N_f 
\right.\nonumber\\
&\left.\quad\qquad +\left(
{\smallfrac {14807}{27648}} 
+{\smallfrac {125}{768}}\, \zeta_3 
-{\smallfrac {5}{36}}\,\zeta_5 \right) N_f^{2} 
\right] N_c^2 \nonumber\\
&\quad
+\left[
\left({\smallfrac{1}{36}} 
-{\smallfrac {13}{48}}\,\zeta_3\right) N_f 
+\left( {\smallfrac {527}{4608}} 
+{\smallfrac {143}{96}}\,\zeta_3 
-{\smallfrac {55}{24}}\,\zeta_5 
\right) C_f^{2} N_f \right.\nonumber\\
&\left.\quad\qquad 
+\left( {\smallfrac {2357}{2304}} 
-{\smallfrac {43}{96}}\,\zeta_3 
-{\smallfrac {5}{24}}\,\zeta_5 \right)
 C_f N_f^{2}\right.\nonumber\\
&\left.\quad\qquad 
-\left( {\smallfrac {7}{648}}
+{\smallfrac {7}{432}}\,\zeta_3\right) N_f^{3} 
\right] N_c\nonumber\\*[0.5ex]
&\quad
+{\smallfrac {23}{256}} C_f^{3} N_f
+\left({\smallfrac {11}{576}} 
-{\smallfrac{1}{24}}\,\zeta_3\right)  N_f^{2}
\nonumber\\
&\quad
-\left( {\smallfrac {29}{1152}} 
+{\smallfrac{1}{3}}\,\zeta_3 
-{\smallfrac {5}{12}}\,\zeta_5 \right) C_f^{2} N_f^{2}
\nonumber\\
&\quad
-\left({\smallfrac{1}{16}} -{\smallfrac{1}{24}}\,\zeta_3
\right) C_f N_f^{3} \;-
{\smallfrac {11}{192}}\,{\smallfrac { N_f^{2}}{ N_c^{2}}}
+{\smallfrac{1}{8}}\,\zeta_3 {\smallfrac { N_f^{2}}{ N_c^{2}}}\nonumber\\
&\quad
+B^{\xi}_{30}+B^{\xi}_{31}N_f+B^{\xi}_{32}N_f^2 + B^{\xi}_{33}N_f^3\;,\nonumber
\end{align}}
\end{subequations}

\vspace{-.4cm}
\noindent
where the $B^{\xi}_{mn}$ all vanish in Landau gauge.\footnote{The
  $N_c=3$ Landau gauge version of these results were first presented
  in Refs.~\cite{vonSmekal:2008ma,Sternbeck:2008au} and subsequently
  confirmed in Ref.~\cite{Boucaud:2008gn}.}
The general expressions for the $B^{\xi}_{mn}$ are rather long and
unilluminating, and hence not included here. The results for the 
phenomenologically most interesting case, $N_c=3$, however, are given in 
\Tab{tab:Nc3Bxi}. 
 
The numerical values of the $\beta_i^{\MM}$'s in Landau gauge are given,
for $N_c=3$ and $N_f=0,\, 2$ and $3$, in \Tab{tab:beta2beta3_MM}.
For the reader's convenience, we give also the 
numerical results for general $\xi$ and $N_f=N_c=3$,
\begin{subequations}
\begin{align}
\beta_0^{\MM} &=2.25,\\
\beta_1^{\MM} &= 4.0- 0.421875\xi -0.28125\xi^{2} + 0.140625\xi^{3},
\\*[1ex]
\beta_2^{\MM} &= 20.9183- 0.552182\xi -0.168435{\xi}^{2}
\\
&\quad
- 0.0187824{\xi}^{3} + 0.171387{\xi}^{4}
- 0.0791016{\xi}^{5},\nonumber\\*[1ex]
\beta_3^{\MM} &= 160.771 + 10.5774\xi -2.46840{\xi}^{2}
\\
&\quad
- 0.145040\xi^{3} + 0.857841\xi^{4}+ 0.245698\xi^{5}
\nonumber\\
&\quad
- 0.113708\xi^{6} + 0.0444946\xi^{7}\, ,\nonumber
\end{align}
\end{subequations}
which results serve to illustrate the weakness of the $\xi$ dependence in the
vicinity of Landau gauge. Note that, while the first coefficient,
$\beta^{\MM}_0$ is gauge independent, and universal, the coefficients
beginning with $\beta_1^{\MM}$ are gauge dependent, as is typical
of momentum subtraction schemes (as usual, the universal value of
$\beta_1^{\MM}$ is obtained only in Landau gauge).

\begin{table}[t]
  \caption{The $B^{\xi}$ terms in Eqs.\eqref{eqs:betaCoeff} for $N_c=3$.} 
  \label{tab:Nc3Bxi}
  \centering\small
  \begin{tabular}{l}\hline
    \rule{0pt}{2.5ex}\!\!
    $B^{\xi}_{10} =-\frac{39}{64}\, \xi -\frac{15}{32}\,\xi^{2} 
      +\frac{9}{64}\,\xi^{3}$, \qquad
    $B^{\xi}_{11}= \frac{1}{16}\,\xi
      +\frac{1}{16}\,{\xi}^{2}$\\*[2ex]
    $B^{\xi}_{20}=\left[ {-\frac {7351}{2048}} 
      +{\frac {891}{512}}\,\zeta_3
      \right] \xi  + \left[ {\frac {2177}{2048}} 
      +{\frac {351}{512}}\,\zeta_3\right]\xi^2$\\ 
    \quad\qquad $-\left[ {\frac {225}{2048}}+{\frac
        {81}{512}}\,\zeta_3\right]\xi^3+ {\frac {531}{2048}}\xi^4
    -{\frac {81}{1024}}\xi^5$ \\*[2ex] 
    $B^{\xi}_{21}= {\frac {101}{384}}\,\xi  -\left[ {\frac
      {59}{1536}}+{\frac {9}{256}}\,\zeta_3
      \right] {\xi}^{2} +{\frac {3}{32}}\,{\xi}^{3}
     -{\frac {15}{512}}\,{\xi}^{4}$\\*[2ex]
    $B^{\xi}_{22} = {\frac {5}{288}}\,\xi
     +{\frac {5}{144}}\,{\xi}^{2}$,\qquad
    $B^{\xi}_{33} = \left[- {\frac{43}{5184}}
       +{\frac {1}{72}}\,\zeta_3\right]
     \xi +{\frac {25}{1728}}\,{\xi}^{2}$\\*[2ex]
    $B^{\xi}_{30} = \left[ {-\frac {3791075}{98304}}
     +{\frac {140271}{2048}}\,\zeta_3
     -{\frac {246915}{8192}}\,\zeta_5  \right] \xi$\\ 
    \quad\qquad 
      $ -
     \left[ {\frac {459983}{32768}}
     -{\frac {31887}{4096}}\,\zeta_3 
     +{\frac {19035}{4096}}\,\zeta_5 \right]\xi^2$\\
     \quad\qquad $-\left[ {\frac {24561}{16384}} 
      -{\frac {3681}{8192}}\,\zeta_3
      +{\frac {2475}{4096}}\,\zeta_5
      \right] {\xi}^{3}$\\ 
    \quad\qquad 
      $+\left[ {\frac {11277} {8192}}
        -{\frac {6327}{4096}}\,\zeta_3
        +{\frac {7155}{8192}}\,\zeta_5
      \right] {\xi}^{4} $ \\
      \quad\qquad $+ 
      \left[ {\frac {2727}{4096}}
        +{\frac {513}{8192}}\,\zeta_3
        -{\frac {945}{8192}}\,\zeta_5  \right] {\xi}^{5}-{\frac
        {2619}{16384}}\,{\xi}^{6} 
      +{\frac {729}{16384}}\,{\xi}^{7}$\\*[2ex]
      $B^{\xi}_{31} = 
      \left[ {\frac {203483}{24576}}
        -{\frac {19821}{2048}}\,\zeta_3
        {\frac {2655}{2048}}\,\zeta_5 \right] \xi +
      \left[ {\frac {38057}{12288}} 
        - {\frac {345}{512}}\,\zeta_3
      \right] {\xi}^{2} $\\
     \quad\qquad $+\left[ {\frac {1569}{2048}}
       -{\frac {39}{128}}\,\zeta_3
       -{\frac {15}{2048}}\,\zeta_5  \right] {\xi}^{3}$\\ 
    \quad\qquad 
      $ +
     \left[ {\frac {81}{256}}
       +{\frac {21}{1024}}\,\zeta_3
       -{\frac {315}{4096}}\,\zeta_5 \right] {\xi}^{4} -
     {\frac {513}{4096}}\,{\xi}^{5}
     +{\frac {63}{4096}}\,{\xi}^{6}$\\*[2ex]
     $B^{\xi}_{32} =\left[-{\frac {11}{36}}
       +{\frac {251}{384}}\,\zeta_3 \right] \xi -
     \left[ {\frac {119}{3072}}
       -{\frac {3}{256}}\,\zeta_3 \right] {\xi}^{2} +
     {\frac {15}{512}}\,{\xi}^{3}
     -{\frac {5}{128}}\,{\xi}^{4}$\\*[1ex]
\hline
  \end{tabular}
\end{table}

\begin{table}
  \centering
  \caption{The $\beta_i^{\MM}$ in Landau gauge
    ($\xi\equiv0$) for different~$N_f$ and $N_c=3$.}
  \label{tab:beta2beta3_MM}\small
  \begin{tabular*}{\linewidth}{@{\extracolsep{\fill}}l@{\qquad}rrr}
    \hline\rule{0pt}{2.5ex} 
    & $N_f=0$ & $N_f=2$ & $N_f=3$\\*[0.2ex]
    \hline\rule{0pt}{2.8ex} 
    $\beta_0^{\MM}$ & 2.75  & $2.41\overline{66}$ & 2.25 \\
    $\beta_1^{\MM}$ & 6.375 & $4.791\overline{66}$ & 4.00 \\
    $\beta_2^{\MM}$ & 47.5075357 & 29.1756592 & 20.9183135 \\
    $\beta_3^{\MM}$ & 392.7385102 & 226.4690053 & 160.7710385\\*[0.5ex]
    \hline
  \end{tabular*}
\end{table}

\section{Comparing perturbative expansions}
\label{sec:applics}

The definition of the running coupling in (\ref{alpha_minimom})
  has been widely used, and continues to be widely used, in
  phenomenological applications of QCD Green's functions within
  non-perturbative
continuum approaches \cite{Alkofer:2000wg,Fischer:2006ub} 
based on DSEs or FRGEs, for example. The precise definition of the
underlying renormalisation scheme, the MiniMOM scheme, puts these
approaches on a firmer ground, and should serve to
resolve any previous misunderstandings. We have already stressed
its utility in providing a route to a lattice determination of
$\alpha_s$ requiring only a calculation of two-point functions, which
are relative easy to determine with high precision in current
simulations. Here we show, as an added bonus, that the MiniMOM
coupling may provide a useful alternative to the $\MSb$ coupling in
certain phenomenological applications. We do so by considering the
expressions for the perturbative contribution to quantities relevant
to a selection of phenomenological applications in the $N_f=3$ regime,
expanded in terms of either the $\MSb$ or the MiniMOM coupling. With
$a_{\MSb}=\alpha_s^{\MSb}(Q^2)/\pi$, $a_{\MM}=\alpha_s^{\MM}(Q^2)/\pi$,
$D_1$, $D_2$ and $D_3$ from Eqs.~(\ref{eqs:Ds}), and
\begin{equation}
  \label{eqs:Ci}
  C_1 = -D_1,\ \ C_2 = -D_2 + 2D_1^2\ \ {\rm and}\
  C_3 = -D_3 + 5D_1 D_2 - 5D_1^3
\end{equation}
an observable, $O$, whose $\MSb$ expansion is
\begin{displaymath}
  O = 1 + A_1\, a_{\MSb} + A_2\, a_{\MSb}^2 + A_3\, a_{\MSb}^3 +
  \ldots,
\end{displaymath}
has an equivalent MiniMOM coupling expansion
\begin{align}
  O = 1 &+ A_1 a_{\MM} +  [A_2 + C_1 A_1]\, a_{\MM}^2\\\nonumber
  &+ [A_3 + 2C_1 A_2 + C_2 A_1]\,a_{\MM}^3 \\\nonumber
  &+ [A_4 + 3C_1 A_3 + (2C_2 + C_1^2)A_2 + C_3 A_1]\,a_{\MM}^4
  + \ldots\;.
\end{align}

In investigating the phenomenological utility
of the MiniMOM coupling, we will consider the $N_c=N_f=3$ case, for which 
$C_1=-2.6875$, $C_2=-2.38108495$ and $C_3=1.3815951$. We show that, in
some of the considered cases, use of the MiniMOM coupling
significantly improves the apparent convergence of the relevant
perturbative series, while, in other cases, it does not. Whether or
not it is useful to employ the MiniMOM coupling is thus something to
be decided on a case to case basis.

\subsection{The Adler function of the vector/axial vector current 
correlators}

Our first example is the dimension \mbox{$D=0$} contribution to
the Adler function, $D_{V/A;ij}(Q^2)\, =\, -Q^2 d\Pi_{V/A;ij} (Q^2)/dQ^2$, 
of the flavour $ij$ vector (V) or axial vector (A) current scalar
correlator, $\Pi_{V/A;ij}$. At scales of phenomenological interest,
$D_{V/A;ij}(Q^2)\big|_{D=0}$ is far and away the dominant term on the OPE
side of finite energy sum rules (FESRs) which have been studied
in the literature based on either 
electroproduction cross-sections or hadronic $\tau$ decay data. 
The $\tau$-based FESRs are used in precision determinations
of $\alpha_s(M_Z)$, the most recent of which
are described in Refs.~\cite{Beneke:2008ad,Maltman:2008nf,Narison:2009vy}.
A combination of electroproduction- and $\tau$-based FESRs has also been used
to investigate the present discrepancy~\cite{Davier:2003pw,Davier:2007ua} 
between the electroproduction and $\tau$ version of the
$I=1$ V spectral function~\cite{Maltman:2005yk}, a discrepancy 
which prevents a clear decision as to whether or not the Standard Model (SM)
prediction for $(g-2)_\mu$ is compatible with the current
high-precision experimental result~\cite{Bennett:2004pv}.

$D_{V/A;ij}\big|_{D=0}$ is known to 5-loops~\cite{Baikov:2008jh} and, 
for \mbox{$N_c=N_f=3$}, given in terms of $a_{\MSb}$, by
\begin{align}
  4\pi^2 D_{V/A;ij}\big|_{D=0} = 1 &+ a_{\MSb} + 1.6398\,
  a_{\MSb}^2 \\\nonumber 
          & + 6.3710\, a_{\MSb}^3 + 49.0757\, a_{\MSb}^4 + \ldots\;.
\end{align}

The equivalent expansion in terms of $a_{\MM}$ is
\begin{align}
\label{adlerMM}
  4\pi^2 D_{V/A;ij}\big|_{D=0} = 1 &+ a_{\MM} - 1.0477\,
  a_{\MM}^2 \\\nonumber
      & - 4.8241\, a_{\MM}^3 + 3.1257\, a_{\MM}^4 + \ldots\;
\end{align}
which displays significantly improved convergence at scales relevant to
the phenomenological studies noted above ($\sim 2-4\ {\rm GeV}^2$),
even when one takes into account the increased size of $\alphaMM(Q^2)$ 
as compared to $\alphaMS(Q^2)$. This improved convergence will also 
be manifest in FESR studies which employ ``contour improved 
perturbation theory'' (CIPT) in their evaluations of the
relevant weighted $D=0$ integrals.\footnote{The CIPT
prescription employs the truncated expansion of Eq.~(\ref{adlerMM})
point-by-point along the circle $\vert Q^2\vert = s_0$ in the
complex $Q^2$-plane. An alternate approach to evaluating the weighted
$D=0$ contour integrals is to use the ``fixed order perturbation theory'' 
(FOPT) prescription, in which the series is expanded, and truncated,
using the running coupling at the same fixed scale, e.g., $\mu^2=s_0$, 
for all points on the circle. In the FOPT scheme, large logs are
unavoidable over some portion of the contour. Nonetheless, recent
arguments~\cite{Beneke:2008ad}, based on a model constrained by
known features of the large order behaviour of the $D=0$ perturbative
series for $D_{V/A;ij}(Q^2)$, shows it is possible that the truncated
FOPT form might provide a more reliable representation of the resummed
series than would the CIPT form. This is potentially relevant here
because the improvement in the convergence of the integrated FOPT
series achieved through the use of the MiniMOM couplant is, for 
commonly used weights, far less compelling than that achieved
in the CIPT case. While a study of the 5-loop FOPT approximation
to a range of weighted integrals of the model for the
resummed series in Ref.~\cite{Beneke:2008ad} shows that a good
representation of the corresponding data integrals is not possible,
in contrast to the situation when the 5-loop CIPT evaluation is
employed~\cite{Maltman:2009ip}, the analogous study has not yet
been performed for the full resummed model, and, as a result,
the preference for the CIPT approach (where the improvement due to
the re-ordering of the series using $a^{\MM}$ would be operative)
is not yet conclusive.}

\subsection{The second derivative of the D=0 part of the
  scalar/ pseudo\-scalar correlator}

As our second example, we consider the
subtraction-constant-independent second derivatives,
$\Pi_{S/PS;ij}^{\prime\prime}(Q^2)$, of the $D=0$ part of the
scalar (S) and pseudoscalar (PS) correlators, 
$\Pi_{S;ij}$ and $\Pi_{PS;ij}$, formed
from the divergences of the flavour $ij$ V or A currents. These
quantities, which are equal for the S and PS cases, apart from the
overall $(m_i\mp m_j)^2$ factors, are the dominant terms on the OPE
side of S and PS FESRs and Borel sum rules (BSRs) which provide the
most reliable sum rule determinations of $m_s$ and
$m_u+m_d$~\cite{Maltman:2001gcMaltman:2001nx,Jamin:2001zrJamin:2006tj,
Chetyrkin:2005kn}. Useful  
{\it lower} bounds on $m_s$ have also been obtained from the PS
$ij=us$ sum rules using a combination of the accurately known $K$ pole
contribution and spectral
positivity~\cite{Becchi:1980vz,Lellouch:1997hp,Chetyrkin:2005kn}. A
combination of FESRs and BSRs based on $\Pi_{PS;us}$, in addition,
provides a determination, not just of $m_s$, but also of the decay
constants of the first two excited $K$ resonances, and hence a useful,
highly constrained model of the $us$ PS spectral function, a model
which, combined with the $us$ S spectral function constructed in
Refs.~\cite{Jamin:2001zrJamin:2006tj},
allows the continuum $J=0$ spectral contributions to be subtracted
from the experimental differential distribution in strange hadronic
$\tau$ decays. This turns out to be a crucial 
input to the hadronic $\tau$ decay determination of 
$\vert V_{us}\vert$~\cite{Gamiz:2002nuGamiz:2004arGamiz:2007qs,
Maltman:2006stMaltman:2007icMaltman:2007pr,Pich:2008ni,Maltman:2008ib}
since the OPE representation of the $J=0$ contributions is
extremely badly behaved at all kinematically accessible scales,
preventing one from employing FESRs based on the full
experimental differential distribution~\cite{Maltman:1998qzMaltman:2001sv}. 

The expansion of $\Pi^{\prime\prime}_{S/PS;ij}$ in terms of 
$a_{\MSb}$ is known to five loops~\cite{Chetyrkin:2005kn} 
and, for $N_c=N_f=3$, is given by 
\begin{align}
& \Pi_{S/PS;ij}^{\prime\prime}(Q^2)\big|_{D=0} \, =\,  
\frac{3[(m_i \mp m_j)(Q^2)]^2}{8\pi^2 Q^2}
          \bigg[1 + \frac{11}{3} a_{\MSb}\; + \\\nonumber
 &+ 14.17928\,a_{\MSb}^2 
   + 77.36834\,a_{\MSb}^3
 + 511.82848\,a_{\MSb}^4 +\ldots\bigg]\, ,
\intertext{where $m_i(Q^2)$ is the running quark mass in the
$\MSb$ scheme. Re-expressing the series in terms of $a_{\MM}$ yields}
&  \Pi_{S/PS;ij}^{\prime\prime}(Q^2)\big|_{D=0} = 
\frac{3[(m_i \mp m_j)(Q^2)]^2}{8\pi^2 Q^2}
           \bigg[1 + \frac{11}{3} a_{\MM} \; + \\\nonumber
&+ 4.32512\, a_{\MM}^2 - 7.57595\, a_{\MM}^3 - 71.99997\, a_{\MM}^4 +
\ldots\bigg]\,,
\end{align}
which again displays significantly improved convergence. Such improved
convergence is likely to allow a significant reduction in
the errors on the determination of the light quark masses, and
an improved version of the light quark mass bounds.

\subsection{The leading $D=2$ contribution to the flavour-breaking
$ud$-$us$, V+A, $J=0+1$ correlator difference}

The flavour-breaking correlator difference, 
$\Delta\Pi_\tau\equiv \Pi^{(0+1)}_{V+A;ud}-\Pi^{(0+1)}_{V+A;us}$, where the 
superscript $(0+1)$ denotes the sum of $J=0$ and $1$ contributions, is of 
interest in extracting $\vert V_{us}\vert$ (and/or $m_s$)
from hadronic $\tau$ decay data. The leading term in the OPE
representation of $\Delta\Pi_\tau$ is the $D=2$ mass-dependent perturbative 
contribution, $\Delta\Pi_\tau\big|^{\mathrm{OPE}}_{D=2}$, proportional to
$m_s^2$. FESRs based on the $J=0+1$ combination are employed because
of the very bad behaviour of the OPE representation of the related
integrated $D=2$, $J=0$
contribution~\cite{Maltman:1998qzMaltman:2001sv}. Even after 
the subtraction of $J=0$ spectral contributions made possible by
the $us$ S and PS studies noted
above~\cite{Maltman:2001gcMaltman:2001nx,
Jamin:2001zrJamin:2006tj}, the
$\vert V_{us}\vert$ (and/or $m_s$) extraction is complicated by
the slow convergence of the $D=2$, $J=0+1$ series at the correlator
level. $\Delta\Pi_\tau\big|^{\mathrm{OPE}}_{D=2}$ is known to 
4-loop order and, for $N_c=N_f=3$, neglecting corrections of 
$O(m_u^2/m_s^2)$, has the form~\cite{Baikov:2004tk}
\begin{align}\nonumber
\label{fbtaumsbar}
  \Delta\Pi_\tau (Q^2)\big|^{\mathrm{OPE}}_{D=2} 
   ~=~& \frac{3 m^2_s(Q^2)}{2\pi^2Q^2}
   \bigg[1 + \frac{7}{3}a_{\MSb} + 19.93313\, a_{\MSb}^2 \\
   & + 208.746\, a_{\MSb}^3 + (2378)\,a_{\MSb}^4 + \ldots\bigg]
\end{align}
where the PMS/FAC estimate for the 5-loop coefficient,
2378~\cite{Baikov:2004tk},  
has been included for exploration purposes. Since $a_{\MSb}(m_\tau^2)\sim 0.1$,
convergence is rather slow at the space-like point on
the FESR contour. It would be very helpful in reducing theoretical 
uncertainties in the determination of $\vert V_{us}\vert$ by this
method were the use of the alternate couplant, $a_{\MM}$, 
to significantly improve the convergence of the $D=2$, $J=0+1$ series. 
Recasting (\ref{fbtaumsbar}) terms of $a_{\MM}$, we find
\begin{align}\nonumber
 & \Delta\Pi_\tau (Q^2)\big|^{\mathrm{OPE}}_{D=2} ~=~ \frac{3
   m^2_s(Q^2)}{2\pi^2Q^2} \bigg[1 + \frac{7}{3}a_{\MM} + 13.66230\, a_{\MM}^2 \\
        &\quad + 96.04956 \, a_{\MM}^3 + (747.25429)\, a_{\MM}^4 
+ \ldots\bigg]\ .
\end{align}
Unfortunately, even at scales $\sim m_\tau^2\simeq 3.16\ {\rm GeV}^2$,
where $ a_{\MM}(m_\tau^2)\sim 0.15  $,
the decrease in the coefficient sizes for this alternate representation
is roughly compensated for by the increased size of $a_{\MM}$ as compared to 
$a_{\MSb}$. Thus, in this case, the MiniMOM coupling does not produce a useful
re-ordering of the original series.

\section{Conclusions}
\label{sec:conclusion}

Our main intention here was to provide the precise definition
  together with a perturbative analysis of the
MiniMOM scheme underlying the running coupling in
(\ref{alpha_minimom}) which has been widely used in non-perturbative
studies of the infrared behaviour of QCD Green's functions and
phenomenological applications since its introduction more than 12 years ago in
Refs.~\cite{vonSmekal:1997isvonSmekal:1997vx}.

The particular occasion for this probably overdue clarification
  is the relatively recent and promising effort to determine this
  coupling from lattice simulations. The high precision and reliable
  error estimates desirable for this project require the improved
  perturbative knowledge which our relation between the MiniMOM and
  the $\MSb$ couplings at 4-loop order provides. When determining  the
  MiniMOM coupling from the gluon and ghost propagators of lattice
  Landau gauge in Monte Carlo simulations, with  
discretisation errors of $O(a^2)$, one has 
\begin{equation}
  \label{eq:running_coup_latt}
  \alphaMM(q^2) = \frac{g^2(a)}{4\pi}\, Z_L(q^2,a^2)\,
  G^2_L(q^2,a^2) + O(a^2)\,,
\end{equation}
as $a\to0$, with $g^2(a)$ the bare coupling at the lattice cutoff scale
$1/a$, and $Z_L$ and $G_L$ are the bare lattice gluon and ghost
dressing functions, respectively (see \cite{Sternbeck:2007br} for
details). It turns out that $\alphaMM$  
can be determined quite accurately over a wide range of scales. 
Once the lattice scale is fixed, this translates into a high-precision
determination of $\Lambda^{\MM}$. $\Lambda^{\MSb}$ can then be
determined via 
\begin{equation}
  \label{eq:LambdaMM_LambdaMS}
  \frac{\Lambda^{\MM}}{\Lambda^{\MSb}}=\exp\left( {\frac{D_1}{2\beta_0}}\right)
\end{equation}
where $D_1$ is given by Eq.~(\ref{ZG2-MSb-expl-1}) and, with our normalisation,
for $N_c=N_f=3$ for example, $\beta_0=9/4$. Numerical values 
for the $N_c=3$, Landau gauge version of this conversion factor are 
given in \Tab{tab:lambdaratios}. A thorough up-to-date analysis
using the available lattice data for $N_f=0,2$ is given
elsewhere \cite{Sternbeck:2009la}.

Finally, as we have demonstrated in the previous section, it turns out
that the MiniMOM coupling is also well suited for use in
perturbative analyses in the place of the commonly used $\MSb$
coupling, since the relation between the two is known to 4-loop order. We
have compared the corresponding perturbative expansions of a selection
of observables of phenomenological interest and found that the MiniMOM
coupling appears to provide an efficient re-ordering of the leading 
$D=0$ perturbative contributions for the V/A and S/PS correlator
cases. One should bear in mind, however, that this improvement
     is not universal, as the example of the $D=2$ contribution to the
     flavour-breaking, $J=0+1$, $V+A$ correlator difference
     shows, though this case does not create a dramatic
     problem either. Those examples where significant improvement is found,
     however, suggest it may be useful to consider $\alphaMM$ as an
     alternative expansion parameter for the perturbative series
     relevant to other observables, such as those relevant to the
     decays of heavy quarkonia and heavy quark physics, as well.

\begin{table}[t]
  \caption{$\Lambda^{\MM}/\Lambda^{\MSb}$ in Landau gauge for 
    $N_c=3$ and different $N_f$.}
  \label{tab:lambdaratios}
  \centering\small
  \begin{tabular*}{\linewidth}{@{\extracolsep{\fill}}lllll}
    \hline\rule{0pt}{2.5ex}
    $N_f=0$ & $N_f=2$ & $N_f=3$ & $N_f=4$ & $N_f=5$\\*[0.5ex]
    \hline \rule{0pt}{2.5ex}
    $1.8968$ &$1.8469$ &$1.8171$ &$1.7831$ & $1.7440$\\*[0.2ex]
    \hline
  \end{tabular*}
\end{table}

\section*{Acknowledgements}

This research was supported by the Australian Research
Council. K.~M. acknowledges the ongoing support of the Natural Sciences
and Engineering Council of Canada, and the hospitality of the CSSM at
the University of Adelaide.



\end{document}